\documentclass[11pt]{article}
\pdfoutput=1
\usepackage{amsmath,amssymb,epsfig,amsfonts}
\usepackage{graphicx,subfigure}
\usepackage[usenames, dvipsnames]{color}
\usepackage[pagebackref,hidelinks]{hyperref}
\usepackage{cite}
\usepackage{verbatim} 
\usepackage{float}
\restylefloat{table}
\usepackage[all]{xy}
\usepackage{pdflscape}
\usepackage{tikz}
\usetikzlibrary{shapes.geometric}
\usepackage{array}
\usepackage{youngtab}
\usepackage{multirow}
\usepackage{color}
\usepackage{ulem}
\usepackage{tikz-cd}
\usepackage{xcolor}
\usepackage{tikz}
\usepackage{tikz-3dplot}

\DeclareFontFamily{U}{wncy}{}
    \DeclareFontShape{U}{wncy}{m}{n}{<->wncyr10}{}
    \DeclareSymbolFont{mcy}{U}{wncy}{m}{n}
    \DeclareMathSymbol{\Sh}{\mathord}{mcy}{"58} 

\makeatletter

\renewcommand*{\backref}[1]{}
\renewcommand*{\backrefalt}[4]{({%
    \ifcase #1 Not cited.%
          \or Page~#2.%
          \else Pages #2.%
    \fi%
    })}

\def\CO{{\mathcal{O}}}

\def\bbP{{\mathbb{P}}}

\def\KB{{\overline{K}_B}}

\makeatother
\hypersetup{
    pdftitle={Towards Realistic Yukawa Hierarchies in F-theory},
    pdfauthor={Mirjam Cvetic, Ling Lin, Muyang Liu, Hao Y. Zhang, Gianluca Zoccarato},
    pdfsubject={F-theory, Yukawa couplings, string phenomenology, GUT-model}
}

\begin{document}

\baselineskip=18pt  
\numberwithin{equation}{section}  



\vspace*{-2cm} 
\begin{flushright}
{\tt UPR-1299-T}\\
\end{flushright}

\vspace*{0.8cm} 
\begin{center}
 {\LARGE Yukawa Hierarchies in Global F-theory Models}

 \vspace*{1.8cm}
 {Mirjam Cveti{\v c}$^{1,2,3}$, Ling Lin$^1$, Muyang Liu$^1$, Hao Y.~Zhang$^1$, Gianluca Zoccarato$^1$}

 \vspace*{1cm} 

{\it $^1$Department of Physics and Astronomy, University of Pennsylvania,  \\
 Philadelphia, PA 19104-6396, USA}

\bigskip
{\it $^2$Department of Mathematics, University of Pennsylvania,  \\
 Philadelphia, PA 19104-6396, USA}

\bigskip
{\it $^2$Center for Applied Mathematics and Theoretical Physics, University of Maribor, \\
 Maribor, Slovenia}

 \bigskip
 {{\tt cvetic@physics.upenn.edu}$\,,\quad$ {\tt lling@physics.upenn.edu}$\,,\quad$ {\tt muyang@sas.upenn.edu}$\,,$ \\
  {\tt zhangphy@sas.upenn.edu}$\,,\quad$ {\tt gzoc@sas.upenn.edu}}
%
%
%
\vspace*{0.8cm}
\end{center}
\vspace*{.5cm}
%
\noindent
We argue that global F-theory compactifications to four dimensions generally exhibit higher rank Yukawa matrices from multiple geometric contributions known as Yukawa points.
The holomorphic couplings furthermore have large hierarchies for generic complex structure \mbox{moduli}.
Unlike local considerations, the compact setup realizes these features all through geometry, and requires no instanton corrections.
As an example, we consider a concrete toy model with $SU(5) \times U(1)$ gauge symmetry.
From the geometry, we find two Yukawa points for the ${\bf 10}_{-2} \, \bar{\bf 5}_6 \, \bar{\bf 5}_{-4}$ coupling, producing a rank two Yukawa matrix.
Our methods allow us to track all complex structure dependencies of the holomorphic couplings and study the ratio numerically.
This reveals hierarchies of ${\cal O}(10^5)$ and larger on a full-dimensional subspace of the moduli space.


\newpage

\tableofcontents

\section{Introduction}

The Standard Model of particle physics has many peculiar features, responsible for the rich phenomenology that ultimately shape our macroscopic universe.
One of these features is the texture of Yukawa couplings between different matter generations, which leads to the observed hierarchy of fermion masses.
To be a viable UV completion of our physical world, it is therefore paramount for string theory to be able to reproduce these textures in compactification scenarios.

A promising regime where we can construct and study globally consistent, four-dimensional (4d) string compactifications is F-theory \cite{Vafa:1996xn}, an extension of weakly coupled type IIB string theory that incorporates non-perturbative back-reactions of 7-branes.
There has been a lot of compact model building efforts in this framework, ranging from supersymmetric GUTs\footnote{For a comprehensive list of the vast number of literature in this direction, we refer to section 10.1 of \cite{Weigand:2018rez} and references therein.} to Pati--Salam models or Standard-Model-like examples \cite{Lin:2014qga,Cvetic:2015txa,Lin:2016vus,Cvetic:2018ryq,Cvetic:2019gnh}.
Along with advancements in our understanding of abelian symmetries in F-theory (see \cite{Cvetic:2018bni} and references therein for a review), these works also developed conceptual and practical tools that give us relatively good control over the chiral spectrum.
More recently, we have also learned about methods that, at least in principle, allow us to determine and engineer light vector-like states for compact models \cite{Bies:2014sra,Bies:2017fam}.

In contrast, explicit computations of Yukawa couplings have only been performed in ultra-local models \cite{Heckman:2008qa,Heckman:2009mn,Hayashi:2009ge,Randall:2009dw,Font:2009gq,Cecotti:2009zf,Conlon:2009qq,Hayashi:2009bt}.
That is, the geometry is restricted to the vicinity of a single point on the 7-brane's internal world-volume, where three matter curves meet.
The coupling is then computed as an overlap of the internal wave functions for the various participating 4d chiral multiplets at such an Yukawa point.
In particular, the calculation always results in a rank one Yukawa matrix despite having multiplet chiral generations in each participating representation.
To enhance the rank for the coupling matrix obtained at one point, one typically has to invoke more subtle structures such as T-branes or Euclidean D3-instantons \cite{Marchesano:2009rz,Cecotti:2010bp,Chiou:2011js,Aparicio:2011jx,Font:2012wq,Font:2013ida,Marchesano:2015dfa,Carta:2015eoh}.
An open question is then how such structures can be consistently realized in global models.

In this work, we argue that an elaborate analysis of these subtle issues become obsolete in global models: contributions to the same coupling from \textit{different} Yukawa points will in general add up to a higher-rank coupling matrix.
In simple terms, this is because the eigenbases of wave functions which diagonalize the Yukawa matrices at different points are in general different.
Said differently, an eigenfunction with eigenvalue 0 of the Yukawa matrix at the first point need not be such a function at the second point.
We stress that this effect is purely a consequence of the global geometry, which may be interpreted as ``non-perturbative'' corrections to the ultra-local results.
Certainly, it further highlights the power of F-theory of geometrizing certain instanton effects \cite{Collinucci:2016hgh}, the latter of which are also vital to enhance the Yukawa rank in perturbative type II compactifications \cite{Cremades:2003qj,Cvetic:2003ch,Cremades:2004wa,Blumenhagen:2006xt, Blumenhagen:2007zk}.

The appearance of multiple Yukawa points and their collected contributions is in general expected for global models \cite{Hayashi:2009bt} (see also \cite{Blumenhagen:2006xt, Blumenhagen:2007zk} for similar effects in type II).
In this work, a similar analysis of the wave function behavior has been presented, showing how their profiles change along matter curves, and can lead to independent Yukawa contributions at different points.
The nevertheless ``local'' analysis showed that the eigenvalues are of same order of magnitude.
However, it does not provide an explicit embedding of these effects, localized on the matter curve, into a global setting.
It is in this embedding where our analysis suggests that, in fact, the couplings \emph{can} generically exhibit hierarchical structures.

For an explicit demonstration of these phenomena, we construct a global toy model with $SU(5) \times U(1)$ gauge symmetry with a $G_4$-flux that induces chiral excesses for ${\bf 10}_{-2}$, $\bar{\bf 5}_6$, and $\bar{\bf 5}_{-4}$ states.
These states are localized on $\mathbb{P}^1$s inside the world-volume of the 7-branes supporting the $SU(5)$.
For these $\mathbb{P}^1$s, we can write down explicitly a basis for the holomorphic wave functions, and compute their overlap at the two distinct points where all three curves meet.
These two contributions facilitate a rank two Yukawa matrix for the ${\bf 10}_{-2} \, \bar{\bf 5}_6 \, \bar{\bf 5}_{-4}$ coupling.
Moreover, we can explicitly track the complex structure dependence of the two independent couplings.
Numeric analysis reveals that there are large hierarchies in generic parts (that is, \textit{not} on a lower-dimensional subspace) of the complex structure moduli space.

For simplicity, we only consider the holomorphic part of the Yukawa coupling which enters the superpotential.
To obtain the physical couplings, one would need to properly take into account the K\"ahler-moduli dependent prefactors which canonically normalizes the wave functions' kinetic terms.
However, these prefactors cannot change the rank of the coupling matrix set by the holomorphic piece.
Also, it would be highly unexpected if these normalizations can erase all ``holomorphic'' hierarchies which are independent of K\"ahler moduli.

The paper is organized as follows.
In section \ref{sec:background}, we review the necessary background material for global 4d F-theory compactifications, focusing on the description of chiral matter.
In particular, we explain in section \ref{sec:gauge_theory_basics} the method to compute the holomorphic couplings via a local, Higgs bundle description of the 7-brane's gauge theory.
We use these two complementary perspectives to construct a toy model with a rank two Yukawa coupling in section \ref{sec:toy_model}.
Specifically, we discuss the complex structure dependence of the hierarchy between the two independent couplings in section \ref{subsec:numerical_analysis}.
We close with some remarks on future directions after a summary in section \ref{sec:conclusions}.

\section{Yukawa Couplings in F-theory}\label{sec:background}

Before we explain the details of the computation of Yukawa couplings in F-theory, we briefly summarize the necessary background material.
For more explanation, we refer to recent reviews \cite{Weigand:2018rez,Cvetic:2018bni}.

The physics of F-theory compactification is encoded in an elliptically fibered Calabi--Yau fourfold $\pi: Y_4 \rightarrow B_3$, where $B_3$ can be viewed as the compactification space of the dual strongly coupled type IIB description.
Over the codimension one locus $\{\Delta = 0\} \equiv \{\Delta\} \subset B_3$ wrapped by 7-branes, the elliptic fiber degenerates, and form (after blowing up the singularities) the affine Dynkin diagram of the corresponding non-abelian gauge group.
For simplicity, we assume that there is one irreducible component $S_\text{GUT} \subset \{\Delta\}$ that carries a non-abelian gauge factor.

Disregarding more subtle geometries which would give rise to discrete abelian symmetries, we consider fibrations with at least one so-called zero-section. 
That is, there is a rational map $s_0 : B_3 \rightarrow Y_4$ satisfying $\pi \circ s_0 = \text{id}_{B_3}$, which marks a special point on each fiber.
Abelian gauge factors in F-theory arise from additional such rational section, independent of $s_0$.
We shall return to such an example later on.

\subsection{Counting charged matter in F-theory}
\label{subsec:counting_matter_geo}

Charged matter states arise over complex curves $C_{\bf R} \subset B_3$, where the residual component of $\{\Delta\}$ intersects the surface $S_\text{GUT}$.
In the F-theory geometry, such intersections are indicated by enhanced singularities of the elliptic fibration along these curves.

While the singularity structure determines the representation ${\bf R}$ of the matter states, the chiral spectrum is induced by a background gauge flux.
Via duality to M-theory, the flux is a four-form $G_4$ on a resolution of the elliptic fourfold, and the chiral spectrum can be computed via intersection theory on the resolved space (see \cite{Weigand:2018rez} and references therein).
However, a refinement of the data is necessary to keep track of the wave functions associated to the massless chiral and anti-chiral multiplets living on the matter curves.
Following a recent proposal \cite{Bies:2014sra,Bies:2017fam} based on type IIB intuition, these massless modes are counted by the cohomologies
\begin{align}\label{eq:cohomologies_counting_vector_spectrum}
\begin{split}
	\text{chiral:} \quad & H^{0} (C_{\bf R}, {\cal L}_{\bf R} \otimes S_{C_{\bf R}}) \, ,\\
	\text{anti-chiral:} \quad & H^1 (C_{\bf R}, {\cal L}_{\bf R} \otimes S_{C_{\bf R}}) \, , 
\end{split}
\end{align}
where ${\cal L}_{\bf R}$ is a line bundle (more generally, a coherent sheaf) on $C_{\bf R}$ extracted from the $G_4$-flux data, and $S_{C_{\bf R}}$ the spin bundle on $C_{\bf R}$.
In general, these cohomologies vary with complex structure moduli, and only the difference $\chi({\bf R}) = h^0({\cal L}_{\bf R} \otimes S_{C_{\bf R}}) - h^1({\cal L}_{\bf R} \otimes S_{C_{\bf R}}) = \int_{C_{\bf R}} c_1({\cal L}_{\bf R})$ counting the chiral excess remains a topological invariant.

The explicit computation of these cohomologies is in general a very difficult task, and requires extensive computing power \cite{Bies:2017fam}.
In particular, these technical difficulties pose a real challenge in constructing F-theory models with realistic vector-like spectra.
Since this is not our main motivation, however, we will focus on constructions where the relevant matter curves have genus 0, i.e., are $\bbP^1$s.
This restriction leads to a significant simplification, as a $\bbP^1$ has no complex structure deformations.
In practice, we recall that any line bundle on $\bbP^1$ is characterized by a single integer $n$, i.e., ${\cal L} \otimes S = \CO(n)$, and
\begin{align}\label{eq:cohomologies_on_P1}
\begin{split}
	& h^0 (\bbP^1, \CO(n)) = \left\{ 
		\begin{array}{ll}
			n+1 , & \text{ if } \, n \geq 0\\
			0 , & \text{ otherwise}
		\end{array}
	\right.\\
	\text{and } \quad & h^1 (\bbP^1, \CO(n)) = h^0( \bbP^1, \CO(- n - 2)) \, .
\end{split}
\end{align}
From this formula, it is evident that the chiral index $\chi = h^0 - h^1$ determines $n$ uniquely.
In turn, $\chi$ can be easily determined via well-known integral formula that can be evaluated on the (resolved) elliptic fourfold $Y_4$.
Note that in particular, we can never have both $h^0$ and $h^1$ be non-zero, and hence there is never any light vector-like pairs on a $\bbP^1$ matter curve.

\subsubsection{Wave functions and holomorphic Yukawa couplings}

To each (anti-)chiral multiplet in representation ${\bf R}$, we can associate a wave function
\begin{align}\label{eq:full_wave_function}
	\Psi = \psi_\text{loc} \times \eta_\text{hol} \, .
\end{align}
Here, the factor $\psi_\text{loc}$ describes the localization of the wave function over the matter curve.
Locally, one usually has $\psi_\text{loc} \sim \exp(- z \bar{z} \, N)$, where $z$ is the local coordinate on $S_\text{GUT}$ transverse to $C_{\bf R}$, and $N$ the flux units on $C_{\bf R}$.
This leads to a Gaussian localization of the wave function around the matter curve.
The second factor $\eta_\text{hol} \equiv \eta$ is a holomorphic section of the corresponding line bundle. 
For chiral multiplets this is the bundle ${\cal L}$, whereas for anti-chiral it is (via Serre-duality) the bundle ${\cal L}^\vee \otimes K_{C}$, where $(\cdot )^\vee$ denotes the dual bundle, and $K_C$ the canonical bundle of $C_{\bf R}$.

Intuitively, one can understand the Yukawa coupling as a result of the overlap of wave functions at the point where three matter curves meet, receiving two contributions from the holomorphic and non-holomorphic factors.
In order to have a higher rank Yukawa matrix, the holomorphic piece must provide this structure in the first place.

To compute the holomorphic Yukawa coupling in a flux background, we have to pick a basis $\eta_i$ of the holomorphic sections of the appropriate bundle.
The holomorphic coupling $W_{ijk}$ is then given by essentially a residue formula of the sections.
The novelty of this work is that we provide an explicit construction where we can evaluate this formula globally, showing that contributions from different Yukawa points in the base generically lead to a higher rank coupling matrix.

The techniques to evaluate each contribution explicitly are based on the local description of the 7-brane gauge dynamics in terms of a Higgs bundle.
In the following, we review this approach and derive the main formula.

\subsection{8d gauge theory and Yukawa couplings}\label{sec:gauge_theory_basics}

The dynamics on the worldvolume of 7-branes wrapping a divisor $S$ is controlled by a supersymmetric 8d Yang--Mills theory. 
The bosonic fields are a gauge field $A$ of a gauge bundle $E$, and a $(2,0)$-form $\Phi$ in the adjoint representation of the gauge algebra. 
The vacuum expectation value of $\Phi$ captures details of the local F-theory geometry close to the divisor $S$ and, in particular, it encodes the locations of localized matter and the couplings among the various matter fields.\footnote{The fact that a $(2,0)$-form describes the normal deformations of the branes is due to the topological twist \cite{Beasley:2008dc}. In the case of branes embedded in a Calabi--Yau threefold $X$, the $(2,0)$-form corresponding to a given normal holomorphic deformation $v \in H^0(S,N_{S/X})$ can be obtained by contracting the Calabi--Yau $(3,0)$-form $\Omega$ with $v$, that is $\Phi_v = \iota_v \Omega.$} 
Specifically, when the rank of $\Phi$ reduces over a complex codimension one sub-variety $\Sigma \subset S$, we find localized fields that are trapped on $\Sigma$.\footnote{In the weak coupling limit (if applicable) this situation corresponds to the intersection of branes and the matter fields come from open strings stretching between the two intersecting stacks of branes.} 
The reduction of the rank of $\Phi$ implies that a larger gauge algebra is preserved over $\Sigma$, a phenomenon that exactly mirrors what happens in the geometry, and the localized matter and its representation under the gauge group can be read off from the enhancement pattern following \cite{Katz:1996xe}. 
Further enhancement of the gauge algebra can occur at points $p \in S$ where triples of curves $\Sigma_i$ intersect. 
This has the effect of producing a triple coupling among the fields hosted on the three matter curves that will produce a Yukawa couplings in the 4d action. 
The pattern of couplings produced will be dictated by the enhanced gauge group $G_{\text{Yuk}}$ at the Yukawa point \cite{Beasley:2008dc}. 
In the following we will provide a quick review of how to perform the computation of said Yukawa couplings. 

\subsubsection{Gauge theory close to Yukawa points}

We start by describing the configuration of the gauge theory in the proximity of a Yukawa point $\mathfrak p_{\text{Yuk}}$. Since we need an enhancement to at least $G_{\text{Yuk}}$ we take a gauge bundle $E$ with gauge algebra $\mathfrak{g}_{\text{Yuk}}$. The vacuum expectation value of $\Phi$ will leave intact only a sub-algebra $\mathfrak{g}$---the physical gauge algebra---at generic points on $S$, with some further enhancements in codimension one where the matter curves are located. The profile of $\Phi$ and the gauge bundle need to satisfy the following BPS equations,
\begin{align}
\bar{\partial}_A \Phi = 0\,,\\
F^{(0,2)} = 0\,,\\
\label{eq:Dt}\omega \wedge F +\frac{1}{2} [\Phi,\Phi^\dag]=0\,,
\end{align}
to ensure that the resulting 4d theory preserves $\mathcal N=1$ supersymmetry. Here $\omega$ is the K\"ahler form on $S$. The first two conditions can be derived from a superpotential
\begin{align}
\label{eq:superP}
W = \int_S \text{Tr} \left(F \wedge \Phi\right)\,,
\end{align}
by taking variations with respect to $\Phi$ and $A$. 
They imply the following conditions: on $S$ the gauge bundle $E$ has to be holomorphic and moreover $\Phi$ is a holomorphic section of $K_S \otimes \text{ad} (E)$ where $K_S$ is the canonical bundle of $S$. The condition \eqref{eq:Dt} ensures the vanishing of the 4d Fayet--Iliopoulos term. 
While solving this condition is in general extremely complicated we will not need it in the following because we will look only at holomorphic couplings, that is the ones appearing in the superpotential. 
We will discuss the necessary steps to obtain the real couplings at the end of this section. 

In the following we will therefore focus solely on holomorphic data and consider equivalence modulo complexified gauge transformations. 
Another important observation is that the holomorphic couplings will not depend on fluxes \cite{Cecotti:2009zf,Cecotti:2010bp}, implying that knowledge of $\Phi$ is actually sufficient to obtain the couplings. 
This is made more explicit in a gauge where $A^{(0,1)} = 0$, usually called holomorphic gauge.\footnote{The disappearance of fluxes in the superpotential can also be understood as follows: local flux densities depend on volume of two cycles due to quantization conditions, that is they depend on K\"ahler moduli. However the superpotential depends only on complex structure moduli, meaning that fluxes cannot appear. Indeed, we will confirm that holomorphic couplings will be sensitive only to the complex structure moduli.}
This greatly simplifies the background equations because now $\Phi$ simply has to be a holomorphic $(2,0)$-form in the adjoint representation, that is its components when expanded in elements of $\mathfrak g_{\text{Yuk}}$ have to be holomorphic functions.\footnote{This may fail at loci where $\Phi$ develops some poles, however we shall not be interested in this case in the following. Note however that sometimes poles might be unavoidable in compact setups \cite{Marchesano:2019azf}.} 

In this scenario modes will correspond to linear fluctuations around a given background, that is we consider perturbations of the form
\begin{align}
\begin{split}
A_{\bar \imath } & \rightarrow A_{\bar \imath} + a_{\bar \imath}\,,\\
\Phi & \rightarrow \Phi + \varphi\,.
\end{split}
\end{align}
Here we have considered only fluctuations $a_{\bar \imath}$ because they are the ones that enter in the superpotential. 
Holomorphic fluctuations that descend to massless 4d fields will correspond to solutions of the linearized BPS equations in holomorphic gauge,
\begin{align}
\label{eq:F01}\bar{\partial} a & = 0\,,\\
\label{eq:F02}\bar{\partial} \varphi & = i [a ,\Phi]\,.
\end{align}
The general solution of the first equation, at least locally, is $a= \bar \partial \xi$, where $\xi$ is a zero-form.
Then we can solve the second one by setting
\begin{align}\label{eq:sol_linear_BPS}
  \varphi = i[\xi, \Phi] + h\,,
\end{align}
where $h$ is a holomorphic $(2,0)$-form. This characterization is however insufficient because it obscures which modes are localized on matter curves. We will now discuss how to determine which modes are localized and use this information to compute their couplings at the Yukawa point.

\subsubsection{Localized modes and Yukawa couplings}

The missing piece in the description of zero modes involves gauge transformations. Namely, it is necessary to consider an equivalence relation on the space of zero modes of the form
\begin{align}\label{eq:gauge_transf_chi}
\begin{split}
a & \sim a + \bar \partial \chi\,,\\
\varphi & \sim \varphi -i [\Phi,\chi]\,.
\end{split}
\end{align}
Here $\chi$ is the parameter of an infinitesimal gauge transformation. 
We can use this to eliminate the holomorphic $(2,0)$-form appearing in the solution \eqref{eq:sol_linear_BPS} for $\varphi$, however this might not be possible at specific loci where the rank of $\Phi$ drops. Said differently, we can write the so-called torsion condition
\begin{align}\label{eq:tors}
\varphi = -i \left[\Phi, \frac{\eta}{f}\right]\,,
\end{align}
where $\eta$ is regular and $f$ is a holomorphic function vanishing on a curve $\Sigma$. This signals that a mode is trapped on $\Sigma$ as its profile cannot be gauged away via a regular gauge transformation. This gives an explicit algorithm to check whether localized modes exist.
Moreover this information is sufficient to compute the Yukawa couplings between the zero modes.
Indeed, plugging the linearized modes in the superpotential \eqref{eq:superP} we find the triple coupling
\begin{align}
\label{eq:tripl}W_\text{Yuk} = -i \int_S \text{Tr}\left(\varphi \wedge a \wedge a \right)\, .
\end{align}

When evaluated on the solutions of \eqref{eq:F01} and \eqref{eq:F02}, the integral quite remarkably localizes at the Yukawa points \cite{Cecotti:2009zf,Cecotti:2010bp}.
Specifically, let us parametrize the zero modes of ${\bf R}_l$-states localized on a curve $\Sigma_{{\bf R}_l}$ by $h_{{\bf R}_l}^{i_l}$ (which determines $(\varphi_{{\bf R}_l}^{i_l}, \eta_{{\bf R}_l}^{i_l})$ by \eqref{eq:sol_linear_BPS} and \eqref{eq:tors}), where the index $i_l$ labels different chiral ``generations''.
Then the contribution to the coupling between three modes coming from a point $p\in \Sigma_1 \cap \Sigma_2 \cap \Sigma_3$ is given by a residue formula,
\begin{align}\label{eq:residue_formula_general}
  W_{i_1 \, i_2 \, i_3}(p) = -i \, \text{Res}_{p} \left[ \frac{\text{Tr} \left( [\eta_{{\bf R}_2}^{i_2},\eta_{{\bf R}_3}^{i_3}] h_{{\bf R}_1}^{i_1} \right) }{f_2 \, f_3}\right]\,.
\end{align}
Note that the value of this formula is invariant under permuting the role of the three modes, i.e., of which of the representations ${\bf R}_l$ we insert the mode $h_{{\bf R}_l}$ (and $\eta_{{\bf R}_l'}$ for the others), whose corresponding $f_l$ does not appear in the denominator.
For definiteness, we have chosen $l=1$ here.

\subsection{Higher rank coupling from multiple Yukawa points}

In the case of a single Yukawa point $p_1$, previous works \cite{Cecotti:2009zf,Heckman:2008qa,Font:2009gq,Conlon:2009qq} have shown that \eqref{eq:residue_formula_general} leads to a rank one coupling.
More precisely, one can pick a basis $h^{i_l}_{{\bf R}_l}$ for the zero modes on all three curves such that
\begin{align}
  W_{i_1\, i_2 \, i_3}(p_1) = 0 \quad \text{unless } \, i_1 = i_2 = i_3 = 1 \, ,
\end{align}
where we have w.l.o.g.~assumed that it is the first basis element $h^{(l)}_1$ on each curve which couples to the others.
In terms of the residue formula \eqref{eq:residue_formula_general}, this means that for all but the one combination, the functions $\text{Tr} ( [\eta_{{\bf R}_2}^{i_2},\eta_{{\bf R}_3}^{i_3}] h_{{\bf R}_1}^{i_1} ) / (f_2 \, f_3)$ are all regular at $p_1$.

If there is a second point $p_2$ where all curves meet, there is no a priori reason why these functions \textit{all} remain regular at $p_2$.
Instead, the generic expectation is that, even though the contribution $W_{i_1\, i_2 \, i_3}(p_2)$ also has rank one, the corresponding zero modes $\tilde{h}^{i_l}_{{\bf R}_l}$ are linear combinations of $h^{i_l}_{{\bf R}_l}$.
Then, one would in general have two linearly independent zero modes, $h^{1}_{{\bf R}_l}$ and $\tilde{h}^{1}_{{\bf R}_l}$, with a non-zero coupling.

To explicitly verify this expectation, we have to track the zero mode basis elements $h^{i_l}_{{\bf R}_l}$, which are holomorphic sections of a line bundle ${\cal L}_l$ on $\Sigma_{{\bf R}_l}$, from one Yukawa point to the other.
In general, this requires us to identify these sections (given in local coordinates on $\Sigma_{{\bf R}_l}$) as elements of the quotient ring
\begin{align}\label{eq:quotient_ring_general}
  \frac{\mathbb{C}[S]}{\langle f \rangle} \, ,
\end{align}
where $\mathbb{C}[S]$ denotes the regular functions on the K\"ahler surface $S$.
This can be most easily done when $\Sigma_{{\bf R}_l} \cong \bbP^1$, and both Yukawa points $p_1,p_2 \in \Sigma_{{\bf R}_l}$ are within a single $\mathbb{C}^2$ patch with coordinates $(x,y)$ on $S$.
In this case, one can use well-known algebra techniques to find a rational parametrization $t \mapsto (x(t), y(t))$ of the curve satisfying $f(x(t),y(t)) = 0$.
Because this is a birational map, we can invert this relation to obtain representations of polynomials in $t$ as elements of \eqref{eq:quotient_ring_general}, which in the local patch can be modeled as
\begin{align}
  \frac{\mathbb{C}[x,y]}{\langle f(x,y) \rangle} \, .
\end{align}

Let us close this part by connecting the localized modes described thus far back to the geometric perspective on the chiral spectrum in section \ref{subsec:counting_matter_geo}.
First, it is obvious to identify the curves $\Sigma_{{\bf R}_l}$ with the matter curves $C_{{\bf R}_l}$.
Secondly, the first Chern-class of the line bundles ${\cal L}_l$ correspond to magnetic fluxes on the 7-brane's world-volume theory which thread the matter curves.
In the global F-theory picture, this flux is induced by a $G_4$-flux, which restricts on the matter curves to precisely the line bundles appearing in \eqref{eq:cohomologies_counting_vector_spectrum}.
For $C_{{\bf R}} \cong \bbP^1$, a basis for the $N+1 = \int_{C_{\bf R}} c_1({\cal L}_{\bf R})$ independent holomorphic sections can be taken to be the polynomials in the local coordinate $t$ up to degree $N$.

\subsection{Beyond holomorphic couplings}

To really compute the values of the physical couplings it is necessary to go beyond the analysis performed so far. 
It is first necessary to obtain the profile of the background values of $A$ and $\Phi$ in a unitary gauge ensuring that the equation \eqref{eq:Dt} is satisfied as well. 
In addition to this the zero modes in this background will need to satisfy the following equations
\begin{align}
\begin{split}
  \bar{\partial}_A a & = 0\,,\\
  \bar{\partial}_A \varphi & = i [a ,\Phi]\,,\\
  \omega \wedge \partial_A a & = \frac{1}{2} [\Phi^\dag ,\varphi]\,.
\end{split}
\end{align}
The triple coupling can again be computed using \eqref{eq:tripl} and this will yield the same result. What remains to be fixed is the overall normalization of the wave functions.
Namely, the norm of the wave functions determines their K\"ahler potential, and to recover the physical couplings it is necessary to normalize the 4d fields so that their kinetic terms are canonical. 
Given the difficulty of determining these terms in a fully fledged, compact model, we shall henceforth focus only on the holomorphic couplings, leaving the computation of physical couplings for future work.

\section{Compact Toy Model with \texorpdfstring{\boldmath{$SU(5) \times U(1)$}}{SU(5)xU(1)} Symmetry}
\label{sec:toy_model}

In this section, we present a toy model that exhibits a higher rank Yukawa matrix.
We follow the procedure outlined in the previous section to explicitly compute the holomorphic coupling in the global model with all relevant complex structure moduli.
In particular, we demonstrate numerically the dependence of the two independent eigenvalues on the moduli, and show they generically differ by orders of magnitude, that is, have a non-trivial hierarchy.

The underlying geometry is based on a so-called factorized Tate model having an $SU(5) \times U(1)$ symmetry.
While the presence of the abelian factor allows for a simple realization of a chiral spectrum, the known local spectral cover description provides the means to exploit the Higgs bundle approach, and interpret the results in the global geometry.

\subsection[Factorized \texorpdfstring{$SU(5)$}{SU(5)} Tate model with genus-0 matter curves]{Factorized \boldmath{$SU(5)$} Tate model with genus-0 matter curves}
\label{subsec:geometricApproach}

First we would like to give the geometric details of our construction. 
We start form the generic $SU(5)$ Tate model and then impose the so-called 3+2 factorization with an additional tuning.
This specialization has the advantage that, when we place the $SU(5)$ symmetry over a divisor $S_\text{GUT} \cong \text{dP}_2$ of the base, we find three genus-0 matter curves intersecting at two different points on $S_\text{GUT}$.
In addition, the 3+2 factorization leads to the presence of a $U(1)$-symmetry, which we can exploit in this context to give us a concrete $G_4$-flux realizing a spectrum compatible with non-trivial Yukawa couplings.

\subsubsection{Factorized \boldmath{$SU(5)$} Tate Model}

Recall that via Tate's algorithm \cite{Bershadsky:1996nh}, the generic fiber of an elliptic fibration $\pi: Y_4 \rightarrow B_3$ with an $I_5$ singularity over $\{w=0\} = \{w\} \subset B_3$ can be embedded in the weighted projective space $\bbP^{231} \ni [x,y,z]$ as a hypersurface,
\begin{equation}\label{eq:Tate_with_SU5}
P_T := -y^2 + x^{3}+a_{1} x y z+a_{2,1} w \, x^{2} z^{2}+a_{3,2} w^2\, y z^{3}+a_{4,3} w^3\, x z^{4}+a_{6,5} w^5 \, z^{6} = 0\, ,
\end{equation}
where $a_{i,k}$ are sections of $\CO_B(\KB - k\,[w])$, with $[w]$ the divisor class of $\{w\}$ and $\KB$ the anti-canonical class of $B_3$.

In order to have a $U(1)$ symmetry, the fibration needs to have an independent rational section.
The idea of so-called factorized Tate models \cite{Mayrhofer:2012zy} to achieve this is to tune the coefficients $a_{i,k}$ such that the intersection of the hypersurface \eqref{eq:Tate_with_SU5} with $y^2 = x^3$ has more than just the zero-section $Z = \{z=0\}$.
That is, the divisor $\{P_T \} \cap \{y^2 - x^3\} \subset Y_4$ factorizes.

More specifically, by introducing $t \equiv \frac{y}{x}$, we demand that
\begin{equation}\label{eq:factorized_Tate_general}
  \left.P_{T}\right|_{t^2 = x} = t^{5} z a_{1}+t^{4} z^{2} a_{2,1} w+t^{3} z^{3} a_{3,2} w^{2}+t^{2} z^{4} a_{4,3} w^{3}+z^{6} a_{6,5} w^{5} \stackrel{!}{=} -z \prod_{i=1}^{n} Y_{i} \, ,
\end{equation}
where, in addition to the universal zero-section factor $z$, we also find other rational solutions $Y_i =0$.
The careful reader will recognize the resemblance of this condition with split spectral covers \cite{Donagi:2009ra, Marsano:2009gv, Marsano:2009wr, Dolan:2011iu}.
Indeed, the original motivation stems from the spectral cover intuition, and we will later make use of the direct connection when we connect the geometric and the gauge theoretic approaches.

For one independent section, we have $n=2$, and there are two inequivalent factorizations:
The polynomials $Y_{1,2}$ either have degrees $(4,1)$ or $(3,2)$ in $t$.
We will focus on the second case, dubbed the 3+2 factorization, where $Y_i$ take the form
\begin{equation}\label{eq:3+2_factors}
Y_{1}= d_{3} t^{3}+d_{2} t^{2} z+d_{1} t z^{2}+d_{0} z^{3} \, , \quad Y_{2}=c_{2} t^{2}+c_{1} t z+c_{0} z^{2} \, .
\end{equation}

Assuming that we are working over a smooth base $B_3$, all coefficients $a_{i,k}$, $c_j$, $d_l$ that appear can be regarded as elements of a unique factorization domain (UFD).
Then, the factorization condition \eqref{eq:factorized_Tate_general} can be solved for the 3+2 factorization generically as \cite{Marsano:2009wr,Mayrhofer:2012zy}
\begin{equation}\label{eq:generic_solution_3+2_factorization}
    \begin{aligned}
      c_{0} & =\alpha \beta, \quad c_{1}=\alpha \delta, \quad d_{0}=\gamma \beta, \quad d_{1}=-\gamma \delta \, , \\
      \text{and} \qquad a_{6, 5} &= \alpha \beta^2 \gamma,\ \ a_{4, 3} = \alpha\beta d_2 + \beta c_2 \gamma - \alpha\delta^2\gamma,\\
      a_{3, 2} & = \alpha\beta d_3 + \alpha d_2 \delta - c_2\delta\gamma,\ \ a_{2, 1} = c_2 d_2 + \alpha \delta d_3, \ \ a_1 = c_2 d_3 \, .
    \end{aligned}
\end{equation}

One can then straightforwardly determine the codimension two fibers hosting matter states (reading off the $U(1)$ charges $q$ is slightly more involved, see \cite{Mayrhofer:2012zy}).
In the 3+2 factorization, we have two $\mathbf{10}_q$ matter curves on $S_\text{GUT} \equiv \{w\}$,
\begin{equation}\label{eq:10_curves_general}
C_{{\bf 10}_{-2}} : \quad d_{3}=0, \quad C_{{\bf 10}_3} : \quad c_{2}=0 \, ,
\end{equation}
and three $\bar{\bf 5}_q$ matter curves on $\{P_i=0\} \cap S_\text{GUT}$ with
\begin{equation}\label{eq:5_curves_general}
\begin{aligned}
  & C_{\bar{\bf 5}_{6}} : \, \delta = 0, \quad C_{\bar{\bf 5}'_{-4}} :  \, \beta d_{3}+d_{2} \delta  =0\, , \\ 
  & C_{\bar{\bf 5}_{1}} : \, \alpha^{2} c_{2} d_{2}^{2}+\alpha^{3} \beta d_{3}^{2}+\alpha^{3} d_{2} d_{3} \delta-2 \alpha c_{2}^{2} d_{2} \gamma-\alpha^{2} c_{2} d_{3} \delta \gamma+c_{2}^{3} \gamma^{2} = 0\, .
\end{aligned}
\end{equation}

We will focus on the coupling ${\bf 10}_{-2} \, \bar{\bf 5}_6 \, \bar{\bf 5}'_{-4}$ realized at the intersection $\delta = 0 = d_3$ on $S_\text{GUT}$.
For an explicit computation of the Yukawa matrix, we would like to find a concrete model where the involved matter curves have genus 0, and have at least two intersection points.

\subsubsection{Explicit construction with \boldmath{$S_\text{GUT} = \text{dP}_2$}}

To keep the construction simple, we sequester the geometric data and study the problem locally on the surface $S_\text{GUT}$.
For this, we first parametrize the sections that appear in the solution \eqref{eq:generic_solution_3+2_factorization} of the 3+2 factorization in terms of their line bundle (or equivalently, divisor) class, when restricted to $S_\text{GUT}$.
That is, in the following, all sections, line bundles and their dual divisor classes that appear are implicitly understood as the pull-backs/restrictions of the global objects, evaluated in the Picard/homology group of $S_\text{GUT}$.
We will use the notation $[s] = c_1({\cal L})$ to denote the divisor class of a holomorphic section $s \in H^0({\cal L})$, and $\overline{\cal K}$ to label the anti-canonical class of $S_\text{GUT}$.

A priori, the split spectral cover has two free parameters.
One is the first Chern class $c_1(N_{S_\text{GUT}})$ of the normal bundle to the GUT surface inside $B_3$, and the other the difference between the classes of the two factors $Y_1$ and $Y_2$, captured by the class of the coefficient $d_3$.
The other coefficients have to satisfy the ``ladder property":
\begin{align}\label{eq:ladder_property}
\begin{split}
    [d_{j}] &= [d_3] + (3-j) \, \overline{\cal K}\quad (j = 0, 1, 2) \, ,\\
    [c_{i}] &= [c_2] + (2-i) \, \overline{\cal K}\quad (i = 0, 1) \, .
\end{split}
\end{align}
The global realization in terms of the factorized Tate model \eqref{eq:generic_solution_3+2_factorization} further introduces one additional free parameter, $[\delta]$.
The classes of the other coefficients $\{\alpha,\beta,\gamma\}$ are then fully determined by \eqref{eq:generic_solution_3+2_factorization}.
For a consistent model, all these classes must be effective or trivial.

To facilitate a concrete global model, we consider an almost Fano threefold base $B_3$ constructed by blowing up along a nodal curve in $\bbP^3$.
After resolving the singularities and performing a flop transition, one obtains a smooth threefold $B_3 = \tilde{X}$.
While we refer to \cite{Marsano:2009ym} for the explicit construction of $\tilde{X}$, we point out that $\tilde{X}$ contains a rigid dP$_2$ surface, which we identify with $S_\text{GUT}$.

On $S_\text{GUT} = \text{dP}_2$, we will denote by $E_i$, $i=1,2$ the classes of the two exceptional curves, and by $H$ the generic hyperplane class of the underlying $\bbP^2$.
These three classes spans the homology lattice with intersection pairing given by
\begin{align}
  H^2 = 1 \, , \quad E_i \cdot E_j = -\delta_{ij} \, , \quad H \cdot E_i = 0 \, .
\end{align}
The anti-canonical class is $\overline{\cal K} = 3H - E_1 - E_2$.
For an irreducible curve to be a smooth $\bbP^1$, the arithmetic genus,
\begin{equation}
  g = 1 + \frac{1}{2}[C] \cdot ([C] - \overline{\cal K}) \, ,
\end{equation}
must vanish.
On a dP$_2$, the following homology classes have smooth irreducible representatives that have genus 0:
\begin{align}
  [C] \in \{H \, , \, 2H-E_1-E_2 \} \, .
\end{align}
While these by far do not exhaust all possibilities, their corresponding polynomials have small number of free parameters, and thus make the subsequent Yukawa computation easier.

Specifically, we recall that the two curves $C_{{\bf 10}_{-2}}$ and $C_{\bar{\bf 5}_{6}}$ involved in the coupling are given by $d_3 =0$ and $\delta = 0$, and the Yukawa points are localized at $d_3 = 0 = \delta$.
Therefore, the assignment
\begin{align}
  [d_3] = 2H-E_1-E_2 \quad \text{and} \quad [\delta] = H  \, , \qquad \text{with } \quad [d_3]\cdot [\delta]=2,
\end{align}
guarantees that, in addition for the classes $[\delta]$ and $[d_3]$ to have genus 0, they also have to intersect twice, giving two independent contributions to the Yukawa matrix.

By the ladder property \eqref{eq:ladder_property}, $[d_2] = [d_3] + \overline{\cal K} = 5H-2E_1-2E_2$, the third matter curve $C_{\bar{\bf 5}'_{-4}} = \{ \beta d_3 + d_2\delta \}$ would have genus $g = 8 >0$.
Therefore, we further specialize the model such that $C_{\bar{\bf 5}'_{-4}}$ factors into two curve, one of which has genus 0 and passes through the two intersection points at $d_3 = 0 = \delta$.
One straightforward possibility that is compatible with the effectiveness of classes is by tuning $d_2 = \beta \, d_2'$,\footnote{Using $[\beta] + [d_3] = [d_2] + [\delta]$, as evident from the equation of $C_{\bar{\bf 5}'_{-4}}$, the class of $d_2'$ is $[d_2] - [\beta] =  [d_3]-[\delta] = H - E_1-E_2$, which is indeed effective.} in which case the curve splits into
\begin{align}
  C_{\bar{\bf 5}'_{-4}} \rightarrow \{ \beta \} \cup \{d_3 + \delta d_2' \} \, .
\end{align}
The factor $C_{\bar{\bf 5}_{-4}} := \{d_3 + \delta d_2'\}$ clearly passes through the point $\delta = 0 =d_3$, and further has class $[d_3] = 2H-E_1-E_2$ and thus genus 0.
Note that the $\bar{\bf 5}$ states over $C_{\bar{\bf 5}_{-4}}$ will again have $U(1)$ charge $-4$, because the fiber structure which determines this charge is inherited from the original $C_{\bar{\bf 5}'_{-4}}$ curve.
One can check that this tuning does not induce any other codimension two enhancements.

The specific embedding of $S_\text{GUT} \cong \text{dP}_2$ into the base threefold $B_3 = \tilde{X}$ has another pleasant feature here, namely, it avoids non-minimal singularities in codimension three of $B_3$ with the above assignment.
To see this, we recall from \cite{Dolan:2011iu,Mayrhofer:2012zy} that these singularities are at the intersections 
\begin{equation}\label{eq:non-minimal_singularities}
  \{c_2\} \cap \{\alpha\}, \quad \{c_2\} \cap \{\beta\} \cap \{\gamma\}, \quad  \{d_2\} \cap \{d_3\} \cap \{\gamma\} \,
,
\end{equation}
on $S_\text{GUT}$.
For generic choices of complex structure, the last two cases are codimension three \textit{on the surface} $S_\text{GUT}$, and hence absent.
As for the first locus, $\{c_2\} \cap \{\alpha\}$, we note that in the $S_\text{GUT} \hookrightarrow \tilde{X}$ embedding, the normal bundle satisfies $c_1(N_{S_\text{GUT}})|_{S_\text{GUT}} = -H$ \cite{Marsano:2009ym}.
By adjunction, we then have 
\begin{align}
  [a_1] = \overline{K}_B|_{S_\text{GUT}} = \overline{\cal K} + c_1(N_{S_\text{GUT}})|_{S_\text{GUT}} = 2H - E_1 - E_2 = [d_3] \, .
\end{align}
Since $a_1 = c_2 d_3$ in the 3+2 factorization, this means that in our particular model, $c_2$ is just a constant and cannot vanish.
Note that this also eliminates the ${\bf 10}_3$ states on $\{c_2\}$, which however is of no consequence to us here.

To summarize our construction, we consider the 3+2 factorized Tate model (i.e., an $SU(5)$ Tate model \eqref{eq:Tate_with_SU5} with specialization \eqref{eq:generic_solution_3+2_factorization}) on the smooth compact base $B_3 = \tilde{X}$ that was constructed explicitly in \cite{Marsano:2009ym}.
We identify the $SU(5)$ divisor $S_\text{GUT}$ with the rigid dP$_2$ surface inside $\tilde{X}$.
We have shown that a further tuning, $d_2 = \beta\,d_2'$, is compatible with the following divisor class assignments of coefficients of the factorized Tate polynomial when restricted to $S_\text{GUT}$:
\begin{align}\label{eq:divisor_classes_tuned_3+2_model}
  \begin{split}
    & [\alpha] = 2H-E_1-E_2 \, , \quad [\beta] = 4H - E_1-E_2  \, , \quad [\gamma] = 7H - 3E_1-3E_2 \, , \quad [\delta] = H \, , \\
    & [c_2] = 0 \, , \quad [d_2] = 5H - 2E_1-2E_2 \, , \quad [d_3] = 2H-E_1-E_2 \quad (\Rightarrow [d_2'] = H -E_1-E_2) \, .
  \end{split}
\end{align}
This defines a global F-theory model with an $SU(5) \times U(1)$ gauge sector and the following representations on the corresponding matter curves:
\begin{align}\label{eq:matter_curves_tuned_3+2_model}
\begin{split}
  & {\bf 10}_{-2} : \, d_3 =0 \, , \quad \bar{\bf 5}_6 : \, \delta = 0 \, , \quad \bar{\bf 5}_{-4} : \, d_3 + \delta\,d_2' = 0 \, , \\
  & \bar{\bf 5}''_{-4} : \, \beta =0 \, , \quad \bar{\bf 5}_{1} : \, \alpha^{2} \beta^2 {d'_{2}}^{2}+\alpha^{3} \beta d_{3}^{2}+\alpha^{3} \beta d'_{2} d_{3} \delta-2 \alpha \beta d'_{2} \gamma-\alpha^{2} d_{3} \delta \gamma + \gamma^{2} = 0 \, .
\end{split}
\end{align}
There are generically no non-minimal singularity enhancements.
Amongst various types of codimension three singularities, we will focus on the locus $\delta = 0 = d_3$ with $I_2^* \cong SO(12)$ enhancement.
There are $[\delta] \cdot [d_3] = 2$ such points realizing the ${\bf 10}_{-2} \, \bar{\bf 5}_6 \, \bar{\bf 5}_{-4}$ coupling, where all the participating curves have genus 0.

\subsubsection{\boldmath{$U(1)$}-flux and massless spectrum}
\label{subsubsec:flux}

To fully specify the matter content of the 4d F-theory compactification, we need to also incorporate a $G_4$-flux $G_4 \in H^{2,2}(Y_4)$.
In general, the construction of such fluxes is straightforward given an explicit resolution of the fourfold.
Knowing the full space of (vertical) $G_4$-fluxes is oftentimes required for phenomenological purposes, i.e., finding solutions with realistic chiral spectra (see, e.g., \cite{Braun:2011zm, Marsano:2011hv, Krause:2011xj,Grimm:2011fx,Krause:2012yh,Braun:2013yti,Cvetic:2013uta,Cvetic:2015txa,Lin:2016vus,Cvetic:2018ryq,Cvetic:2019gnh}).
Since this is not the primary focus on the present work, our requirements are less constraining.

In particular, it suffices for our purpose to find a flux such that the chiral index for the three representations in question, ${\bf 10}_{-2}$, $\bar{\bf 5}_6$ and $\bar{\bf 5}_{-4}$, are all positive or negative.
This is required as, by gauge invariance and holomorphy of the superpotential, all participating matter fields must be either chiral or anti-chiral.
Since the matter curves are $\bbP^1$s, the number of chiral superfields is the same as the chiral index, as there cannot be any light vector-like pairs.

This condition can be satisfied by just turning on the so-called $U(1)$-flux \cite{Grimm:2010ez,Krause:2011xj},
\begin{align}\label{eq:U1_flux}
  G_4 = \sigma \wedge \omega_B \, .
\end{align}
Here, $\omega_B$ is a $(1,1)$-form that is Poincar\'{e}-dual to a vertical divisor $\pi^{-1}(D_B)$ with $D_B \in H_4(B_3)$.
On the other hand, $\sigma$ is the $(1,1)$-form dual to the so-called Shioda divisor \cite{Park:2011ji, Morrison:2012ei} associated to the rational section generating the $U(1)$.
While we omit the specifics of the construction of this divisor (we instead refer to the reviews \cite{Weigand:2018rez, Cvetic:2018bni} for more details), we note that the $U(1)$-flux \eqref{eq:U1_flux} induces a chirality of a representation ${\bf R}_q$ over a curve $C_{{\bf R}_q}$ given by
\begin{align}\label{eq:chirality_U1_flux}
  \chi({\bf R}_q) = q \, D_B \cdot [C_{{\bf R}_q}] \, ,
\end{align}
where $\cdot$ denotes the intersection product on the base $B_3$.

For the matter states localized on the surface $S_\text{GUT}$, it suffices to parametrize $D_B$ in terms of its restriction to the surface, i.e., $D_B|_{S} = a_1\,E_1 +a_2\, E_2 + b\,H$ with integers $a_1,a_2,b$.
Given the matter curves \eqref{eq:matter_curves_tuned_3+2_model} and their classes, we obtain the following chiralities:
\begin{equation}
  \begin{split}
  \chi({\bf 10}_{-2}) & = -2 \, D_B \cdot [d_3] = -4b-2a_1-2a_2,\quad \chi ({\bar{\bf 5}_6}) = 6\,D_B \cdot [\delta] = 6b,\\
   \chi({\bar{\bf 5}_{-4}}) & = -4 \, D_B \cdot [d_3] = -8b-4a_1-4a_2 .
\end{split}
\end{equation}
Thus, if $\{b>0, (a_1+a_2)<-2b\}$, then all chiralities are positive
For concreteness, we take $(a_1,a_2, b) = (-2,-1, 1)$ to get:
\begin{equation}\label{eq:chiralities}
  \chi({\bf 10}_{-2}) = 2 \, ,\quad \chi ({\bar{\bf 5}_6}) = 6 \, ,\quad \chi({\bar{\bf 5}_{-4}}) = 4 \, .
\end{equation}

\subsection{Higgs bundle description}
\label{subsec:gaugeTheory}

Having laid out the underlying geometric construction, we will now focus on the explicit computation of the Yukawa couplings from the local gauge theory perspective.
As described in section \ref{sec:gauge_theory_basics}, this amounts to viewing the 7-brane's world volume theory on $S_\text{GUT}$ as a Higgs bundle with Lie group $G$.
The $SU(5)$ gauge symmetry in 4d is a result of a breaking by a $G$-adjoint valued Higgs field $\Phi$.

In our toy model, a natural choice for $G$ is provided by the split spectral cover description of the factorized Tate models \cite{Donagi:2009ra,Marsano:2009gv,Marsano:2009wr}.
Motivated by heterotic/F-theory duality, the spectral cover provides a dictionary between the singularity structure over $S_\text{GUT}$ and the localized 8d gauge theory with $G = E_8$.
For the 3+2 factorization, the vacuum expectation value (vev) of $\Phi$ is only non-trivial in $\left(SU(3) \times SU(2) \right)_\perp \times U(1) \subset SU(5)_\perp$, such that the 4d gauge symmetry is the commutant $SU(5)_\text{GUT} \times U(1)$.

We start by decomposing the adjoint $\mathbf{248}$ of $E_8$ to $SU(5)_\text{GUT}\times U(1) \times (SU(3)\times SU(2))_\perp$ and identify the $(SU(3)\times SU(2))_\perp$ representations for the localized fields on each matter curve.
Then we determine how the $(SU(3) \times SU(2))_\perp$-valued $\Phi$ acts on these modes, and solve for the torsion equations \eqref{eq:tors} to obtain the localized wave function $\eta$.

\subsubsection{Localized matter fields from \boldmath{$E_8$} breaking}

The Higgs vev that we impose takes value in $SU(5)_\perp \subset E_8$ and is a field configuration without a T-brane component,
\begin{equation}\label{eq:Higgs_vev}
\Phi = \left(
\begin{array}{ccc|cc}
0  &  0  & -D_3  & 0  &  0  \\
1  &  0  & -D_2  & 0  &  0  \\
0  &  1  &  -D_1 & 0  &  0  \\\hline
0  &  0  &   0     &  0 & -C_2\\
0  &  0  &   0     &  1 & -C_1
\end{array}
\right) \, .
\end{equation}
This Higgs background can be derived from a spectral cover associated with \cite{Marsano:2009gv,Marsano:2009wr}
\begin{equation}
  P = (s^3 + D_1 s^2 + D_2 s + D_3)(s^2 + C_1 s + C_2) \, .
\end{equation}
The obvious connection to the 3+2 factorization \eqref{eq:3+2_factors} of the Tate model is to identify
\begin{equation}\label{eq:identification_local_global}
  D_i = \frac{d_i}{d_0}, \ \ C_i = \frac{c_i}{c_0} \, .
\end{equation}
The locus where this identification breaks down, at $b_0\,c_0 = a_{6,5} = 0$, appear as poles of $\Phi$ in the local description, but are associated to the ``infinity locus" on the spectral cover \cite{Donagi:2009ra}, where one has to move to another patch of the global model.
Since our Yukawa points are not in the vicinity of this locus (see below), we do not have to concern ourselves with this issue.

The 3+2 factorization of the spectral cover implies a further decomposition $SU(5)_\perp \rightarrow (SU(3) \times SU(2))_\perp \times U(1)$, where the $U(1)$ with the generator
\begin{equation}
  \text{diag} \left[ \frac{1}{3}, \frac{1}{3}, \frac{1}{3}, -\frac{1}{2}, -\frac{1}{2} \right] \in SU(5)_\perp
\end{equation}
is unbroken in 4d.
Thus, the Higgs vev \eqref{eq:Higgs_vev} takes value in $(SU(3) \times SU(2))_\perp \times U(1)$ as
\begin{equation}\label{eq:decomposed_higgs}
\Phi = \left(
\begin{array}{ccc}
\frac{D_1}{3} &  0   &   -D_3  \\
1    &   \frac{D_1}{3}    &   D_3  \\
0     &      1     &   -\frac{2D_1}{3}
\end{array}
\right)\oplus \left(
\begin{array}{ccc}
\frac{C_1}{2}  &   -C_2   \\
1       &  -\frac{C_1}{2}
\end{array}
\right)\oplus\left(\frac{D_1}{2}\right) \, .
\end{equation}

In the presence of this vev, the ${\bf 248}$ representation of $E_8$ decomposes into representations $SU(5) \times (SU(3) \times SU(2))_\perp \times U(1) $ as
\begin{equation}
\begin{aligned}
  \mathbf{248} \rightarrow \, & ({\bf 24,1,1})_0 +({\bf 1, 8, 1})_0 + ({\bf 1,1,3})_0 \\
  + & \left( (\mathbf{\bar{5}}, \mathbf{1}, \mathbf{1})_6 + (\mathbf{\bar{5}}, \mathbf{\bar{3}}, \mathbf{1})_{-4} +  (\mathbf{\bar{5}}, \mathbf{3}, \mathbf{2})_{1} + \text{c.c.} \right) + \left( (\mathbf{10}, \mathbf{3}, \mathbf{1})_{-2} + (\mathbf{10}, \mathbf{1}, \mathbf{2})_{3} + \text{c.c.} \right).
\end{aligned}
\end{equation}
In this decomposition, the matter states we are interested in (${\bf 10}_{-2}$, $\bar{\bf 5}_6$ and $\bar{\bf 5}_{-4}$) carry no $SU(2)_\perp$ charges. Still, it is convenient to work with a basis $e_i$ of the fundamental representation of $SU(5)_\perp$ chosen such that the $SU(3)_\perp \subset SU(5)_\perp$ acts only on the elements $e_i$ with $i=1,2,3$. Fields in the anti-symmetric representation of $SU(5)_\perp$, that is the $\mathbf {10}$, can be represented as $e_i \wedge e_j$. With these conventions the matter fields that will be involved in the Yukawa coupling of interested may be represented as
\begin{equation}
\mathbf{10}_{-2} : \left(\begin{array}{c}e_1 \\ e_2\\ e_3\end{array}\right), \quad
\mathbf{\bar{5}}_6 : \left(e_4 \wedge e_5\right), \quad
\mathbf{\bar{5}}_{-4} : \left(\begin{array}{c}e_2\wedge e_3\\e_3\wedge e_1\\  e_1\wedge e_2\end{array}\right) \, .
\end{equation}

The relevance of this basis is that it allows us to readily reconstruct the action of the background Higgs field $\Phi$ once we know its action on the fundamental representation. As an example that will be of interest in the following, the action on the anti-symmetric representation can be obtained as $\Phi(e_i \wedge e_j) = (\Phi e_i) \wedge e_j + e_i \wedge (\Phi e_j)$. 

Now we solve for matter curves and torsion equation for the three matter curves respectively, following section \ref{sec:gauge_theory_basics} (see also \cite{Cecotti:2010bp,Chiou:2011js}).
We also identify the local spectral cover coefficients $C_i = c_i/c_0$ and $D_j = d_j/d_0$ with the global parameters of the factorized Tate model \eqref{eq:generic_solution_3+2_factorization} including the tuning $d_2 = d_2' \beta$.

\begin{itemize}

\item We begin with $\mathbf{\bar{5}}_6 \equiv (\mathbf{\bar{5}}, \mathbf{1}, \mathbf{1})_6$, where we have indicated explicitly the representations under $(SU(3) \times SU(2))_\perp \times U(1) \subset SU(5)_\perp$.
The action of $\Phi$ \eqref{eq:decomposed_higgs} on this field only corresponds to a multiplication by $D_1$. 
Therefore, the rank of $\Phi$ drops along the matter curve
\begin{align}
  \mathbf{\bar{5}}_6: \quad f_{\bar{\bf 5}_6} = D_1 = \frac{d_1}{d_0} = - \frac{\delta}{\beta} = 0\,.
\end{align}
 By using gauge transformation we can fix
\begin{align}
\varphi_{\mathbf{\bar 5}_6} = h_{\mathbf{\bar 5}_6}\,, \qquad h_{\mathbf{\bar 5}_6} \in \frac{\mathbb C[S_\text{GUT}]}{\langle D_1\rangle} \cong \frac{C[S_\text{GUT}]}{\langle \delta \rangle} \, ,
\end{align}
where for the last equality we have used that $\beta$ is a regular function on $S_\text{GUT}$.
Finally we can determine the solution to the torsion equation 
\begin{equation}
 \eta_{\mathbf{\bar{5}}_6} = h_{\mathbf{\bar{5}}_6}\,.
\label{eqn:eta1}
\end{equation}

\item We proceed to $\mathbf{10}_{-2} \equiv (\mathbf{10}, \mathbf{3}, \mathbf{1})_{-2}$, which carries the fundamental representation under $SU(3)$ and two units of $U(1)$ charge. 
To understand how many localized modes we find from this sector, we first perform a general gauge transformation \eqref{eq:gauge_transf_chi} with gauge parameter $\chi$:
\begin{equation}
\delta \varphi_{\mathbf{10}_{-2}} = \left(
\begin{array}{ccc}
0  &  0  & -D_3   \\
1  &  0  & -D_2   \\
0  &  1   & -D_1 
\end{array}
\right)\left(
\begin{array}{c}
\chi_1\\
\chi_2\\
\chi_3
\end{array}
\right)\,.
\end{equation}
The matter curve will be located at the locus where said gauge transformations fail to be invertible as the information contained there cannot be erased via a gauge transformation.  
This happens where the determinant of the matrix vanishes, which in our case is
\begin{equation}
\mathbf{10}_{-2}: \quad f_{{\bf 10}_{-2}} = D_3 = \frac{d_3}{\gamma \beta} = 0\,.
\end{equation}
By using gauge transformations we can gauge away all components of $\varphi_{\mathbf{10}_{-2}}$ but the first one. 
That is, we can choose the gauge to be
\begin{align}
\varphi_{\mathbf{10}_{-2}} = \left(\begin{array}{c} h_{\mathbf{10}_{-2}} \\ 0 \\0 \end{array}\right)\,, \qquad h_{\mathbf{10}_{-2}} \in \frac{\mathbb C[S_\text{GUT}]}{\langle D_3\rangle} \cong \frac{\mathbb{C}[S_\text{GUT}]}{\langle d_3 \rangle} \,.
\end{align}
The solution of the torsion equation is then
\begin{equation}
\eta_{\mathbf{10}_{-2}} = 
\left(
\begin{array}{ccc}
D_2  & -D_3  & 0   \\
D_1  &  0    &  -D_3   \\
1    &  0    &  0   
\end{array}
\right)\left(
\begin{array}{c}
h_{\mathbf{10}_{-2}}\\
0\\
0
\end{array}
\right)
 = 
 \left(
\begin{array}{c}
D_2 \,h_{\mathbf{10}_{-2}}\\
D_1 \,h_{\mathbf{10}_{-2}}\\
h_{\mathbf{10}_{-2}}
\end{array}
\right)\,.
\label{eqn:eta2}
\end{equation}

\item Much of the analysis that we just presented carries out mutatis mutandis for the last matter field, that is the $\mathbf{\bar 5}_{-4}$. We can reconstruct the action of gauge transformation on this field by simply exploiting the fact that it is in the $\mathbf{10}$ representation of $SU(5)_\perp$, obtaining the gauge variation
\begin{equation}
\delta \varphi_{\mathbf{\bar{5}}_{-4}} = 
\left(
\begin{array}{ccc}
-D_1  &  -1  & 0   \\
0  &  -D_1  & -1   \\
D_3  &  D_2  &  0
\end{array}
\right)\left(
\begin{array}{c}
\chi_1\\
\chi_2\\
\chi_3
\end{array}
\right)\,.
\end{equation}
The matter curve is again given by the vanishing locus of the determinant of this matrix action:
\begin{equation}
  \mathbf{\bar{5}}_{-4}: \quad f_{\bar{\bf 5}_{-4}} = D_3 - D_1 D_2 = \frac{d_3 + d_2' \delta}{\gamma\beta} = 0 \, .
\end{equation}
Note how the spectral cover description misses the $\bar{\bf 5}''_{-4}$ curve on $\{\beta\}$ which was a result of our additional tuning.
The fact that we do not see it in the local patch is related to the fact that $\{\beta\}$ is actually part of the ``infinity locus'' of the spectral surface mentioned above.
Again, since this matter curve plays no role in our discussion, we will ignore this issue for the rest of this work. It is possible now to gauge fix $\varphi_{\mathbf{\bar {5}}_{-4}}$ as follows
\begin{align}
\varphi_{\mathbf{\bar 5}_{-4}} = \left(\begin{array}{c}0\\0\\ h_{\mathbf{\bar 5}_{-4}}\end{array}\right)\,,\qquad h_{\mathbf{\bar 5}_{-4}} \in \frac{\mathbb C[S_\text{GUT}]}{\langle D_3-D_1 D_2\rangle} = \frac{\mathbb{C}[S_\text{GUT}]}{\langle d_3 + d_2' \delta \rangle} \,.
\end{align}
Then we have
\begin{equation}
\eta_{\mathbf{\bar{5}}_{-4}} = 
\left(
\begin{array}{ccc}
D_2  &  0  &  1  \\
-D_3  &  0  & -D_1  \\
D_1 D_3 & D_1 D_2 - D_3 &  D_1^2   
\end{array}
\right)\left(
\begin{array}{c}
0\\
0\\
h_{\mathbf{\bar{5}}_{-4}}
\end{array}
\right)
 = 
 \left(
\begin{array}{c}
h_{\mathbf{\bar{5}}_{-4}}\\
-D_1 \,h_{\mathbf{\bar{5}}_{-4}}\\
D_1^2 \,h_{\mathbf{\bar{5}}_{-4}}
\end{array}
\right)\,.
\label{eqn:eta3}
\end{equation}

\end{itemize}

\subsection{Explicit computation of Yukawa couplings}

We now turn to the explicit computation of the Yukawa couplings.
With the above results, we can simply plug in the expressions for $\varphi_{{\bf R}_l}$ and $\eta_{{\bf R}_l}$ into the general formula \eqref{eq:residue_formula_general}.
Assuming we have the different chiral generations represented by $h^i_{{\bf R}_l}$, the triple couplings become
\begin{align}\label{eq:yukawa_final_formula}
  W_{\bar{\bf 5}^i_{-4} \times \bar{\bf 5}^j_{6} \times {\bf 10}^k_{-2}} =   \sum_{P \in \{P_1, P_2\}} \text{Res}_{P} \left[ \frac{\text{Tr}\left( \varphi^i_{\bar{\bf 5}_{-4}} \, \left[ \eta^j_{\bar{\bf 5}_6} , \eta^k_{{\bf 10}_{-2}} \right] \right)}{f_{\mathbf{10}_{-2}}f_{\mathbf{\bar 5}_{6}}} \right] = \sum_{P \in \{P_1, P_2\}} \text{Res}_{P} \left[ \frac{ h^i_{\bar{\bf 5}_{-4}} \, h^j_{\bar{\bf 5}_6} \, h^k_{{\bf 10}_{-2}}  }{ D_1 \, D_3} \right] \, ,
\end{align}
where we have summed up the contributions from the two different points $P_{1,2}$ where this coupling occurs.
In the global model \eqref{eq:generic_solution_3+2_factorization}, the denominator becomes $D_1 \, D_3 = \frac{d_1 \, d_3}{d_0^2} = -\frac{d_3\,\delta}{\gamma \, \beta^2}$.
As we will see, the value of $\gamma \, \beta^2$ can vary by orders of magnitude between $P_1$ and $P_2$, and will be ultimately responsible for large hierarchies between the physical couplings.

The explicit numerical values depend on the complex structure moduli.
For the couplings \eqref{eq:yukawa_final_formula} these are the coefficients of the polynomials of $d_3$, $\delta$, $\beta$ and $\gamma$.
Let us first look at the polynomials $d_3$ and $\delta$, since their mutual vanishing define the points $P_{1,2}$.
In terms of the toric coordinates $(u,v,w,e_1,e_2)$ on $S_\text{GUT} = \text{dP}_2$ (see appendix \ref{app:toric_dP2} for the conventions), the generic polynomials with coefficients $k_i$ satisfying $[d_3] = 2H - E_1 - E_2$ and $[\delta]= H$ are:
\begin{align}\label{eq:explicit_param_d3delta}
\begin{split}
  \delta & = k_0 \, (u e_1 e_2) + k_1 \, (v e_2) + k_2 \, (w e_1) \, ,\\
  d_3 & = k_3 \, (u^2 e_1 e_2) + k_4\, (v w) + k_5 \, (u v e_2) + k_6 \, (u w e_1) \, .
\end{split}
\end{align}
Note that an overall scaling of either polynomials neither change the curve their vanishing defines, nor do they affect ratios of the Yukawa couplings.
That is, there are 5 independent complex structure parameters in $d_3$ and $\delta$.
One can easily show that both solutions of $\delta =0 = d_3$ lie in the local patch $\{u \neq 0, v\neq 0, w\neq 0\}$ (see appendix \ref{app:toric_dP2}).
Using the three independent scaling relations on dP$_2$, we can then set these coordinates to 1 in the patch.
Then $\delta$ and $d_3$ are just polynomials in $e_{1,2}$.
Solving for $\delta=0=d_3$, we indeed find two distinct solutions whose explicit expressions we defer to the appendix, see \eqref{eq:solutions_yukawa_points}.

The last remaining piece of information concerns the choice of the sections $h^i_{\mathbf R}$ appearing in the residue formula. 
The choice of representatives will be specified by the line bundle ${\mathcal L}_{\mathbf R} \otimes S_{C_{\bf R}}$ on each matter curve, and given that all matter curves are genus zero curves this bundle is entirely specified by its first Chern class.
More concretely, if ${\mathcal L}_{\bf R} \otimes S_{C_{\bf R}} = \mathcal O_{\mathbb P^1}(N+1)$ then its holomorphic sections can be chosen to be homogeneous polynomials of degree $N$ in the two projective coordinates of the $\mathbb P^1$, or in a local patch, they will be represented by polynomials of degree up to $N$ in the inhomogeneous coordinate $t$ of the patch.

The $\bbP^1$-coordinate $t$ is related to the (local) surface coordinates $e_{1,2}$ by a birational map.
Given the explicit equations \eqref{eq:explicit_param_d3delta}, determining such a rational map is a basic algebra exercise.
For example, the curve $\{d_3\}$ hosting the fields in the $\mathbf{10}_{-2}$ representation can be parametrized as
\begin{align}
  t \mapsto (e_1(t) , e_2(t)) = \left(\frac{k_5 k_6 - k_5 t - k_3 k_4}{k_3 t }\,, \, \frac{t-k_6}{k_3} \right) \,.
\end{align}
From this, we can represent the sections $h_{\mathbf{10}_{-2}}$ which ought to be polynomials in $t$ as functions in $e_i$.
One canonical choice here would be
\begin{align}
  h_{\mathbf{10}_{-2}} &\in \mathbb C \left[e_2 k_3 + k_6\right] \cong \mathbb{C} [e_2]\,.
\end{align}
A similar computation for the $\bar{\bf 5}_6$ curve $\{\delta\}$ leads to the representation
\begin{align}
  h_{\bar{\mathbf 5}_6} &\in \mathbb C\left[e_2 k_0+k_2\right] \cong \mathbb{C} [e_2]\,.
\end{align}
Note that we could have also written them in terms of $e_1$ by inverting the relation between $e_1$ and $t$. 
This would not have affected our result because the two choices are identical when restricted to the matter curve, i.e., as elements in $\mathbb{C}[e_1,e_2]/\langle f \rangle$ for $f = d_3, \delta$.

One can equally obtain a parametrization for the $\bar{\bf 5}_{-4}$ curve, which however is slightly more cumbersome because of the more complicated expression.
For the purpose of exhibiting the higher rank Yukawa structure, we will fix one chiral ``generation'' of this matter, and compute the coupling matrix $W_{ij}$ for different generations $h^i_{{\bf 10}_{-2}}$ and $h^j_{\bar{\bf 5}_6}$.
For the residue formula, we then simply insert for $h^i_{\bar{\bf 5}_{-4}}$ a constant, which is always a valid basis element for holomorphic sections (unless there are none).\footnote{We are basically treating the ${\bar{\bf 5}_{-4}}$ fields as the Higgs representation, for which there is only one chiral superfield in the ``real'' world.
The triple couplings then form an honest matrix $W_{ij}$, where $(i,j)$ run over the ``quarks/leptons''.
While our chiral spectrum is not realistic as we have multiple Higgs fields ${\bar{\bf 5}_{-4}}$ (which might be remedied in future work with a different $G_4$-flux), we note that, at the level of representations, ${\bar{\bf 5}_{-4}}$ can be actually identified with a Higgs field in an $SU(5)$-GUT theory, where the $U(1)$ is of Peccei--Quinn type \cite{Marsano:2009wr}.
} 
Then, \eqref{eq:yukawa_final_formula} reduces to
\begin{align}\label{eq:yukawa_final_final}
  W_{ij} = \sum_{P \in \{P_1, P_2\}} - \text{Res}_P \left[ \frac{\gamma \, \beta^2 \, h^i_{{\bf 10}_{-2}} \, h^j_{\bar{\bf 5}_{6}} }{d_3\,\delta} \right] \, .
\end{align}
For the chiral spectrum \eqref{eq:chiralities} we can pick the basis $h^i_{{\bf 10}_{-2}} \in \{ 1, e_2\}$, and $h^j_{\bar{\bf 5}_{6}} \in \{1,e_2,e_2^2,..., e_2^5\}$.

What remains is to parametrize the functions $\beta$ and $\gamma$.
Since their divisor classes (see \eqref{eq:divisor_classes_tuned_3+2_model}) are quite large, their explicit polynomial expression is lengthy.
Specifically, $\beta$ has 12 independent parameters, and $\gamma$ has 23.
The concrete value of \eqref{eq:yukawa_final_final} depend on all $5+12+23 = 40$ complex structure parameters that appear in $(d_3, \delta, \beta, \gamma)$.
For generic values, we confirm numerically that $W_{ij}$ is indeed a rank two matrix.

\subsection{Complex structure dependence and Yukawa hierarchies}
\label{subsec:numerical_analysis}

Given the explicit parametrization of the complex structure dependence of this Yukawa matrix, we can give a qualitative analysis of how the holomorphic couplings vary over the moduli space.
To do so, we first pick two of the six generators $h^j_{\bar{\bf 5}_{6}}$, say $1$ and $e_2$.
For generic complex structure, the resulting $2 \times 2$ matrix is still rank two, confirming our expectation that the contributions from two Yukawa points are indeed linearly independent.
The two eigenvalues $\lambda_{1,2}$ of this $2 \times 2$ matrix are the two independent holomorphic Yukawa couplings.

\begin{figure}[!p]
\centering
\subfigure[]{\label{subfig:ratio1}\includegraphics[width=.47\textwidth]{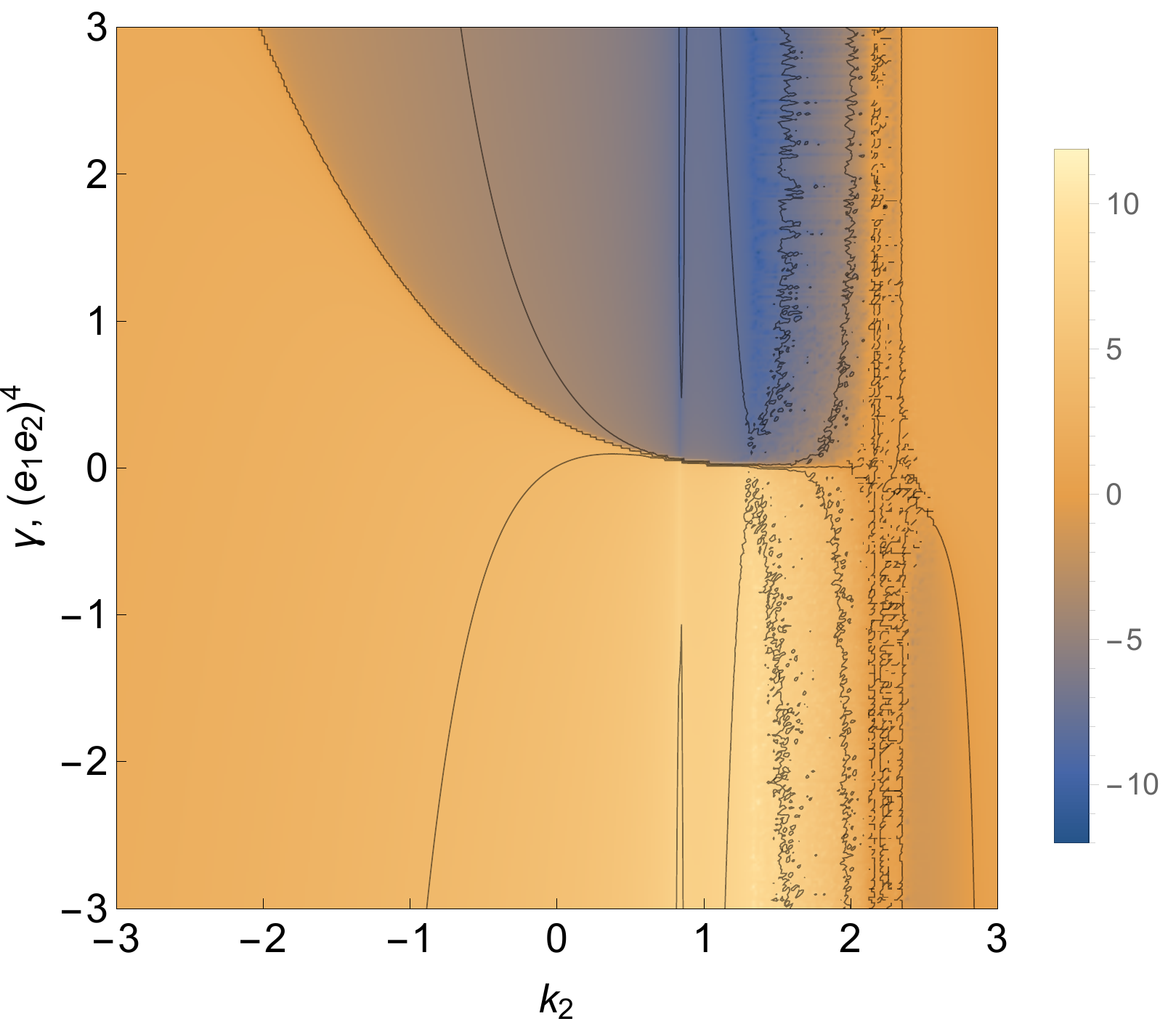}}
\subfigure[]{\label{subfig:ratio2}\includegraphics[width=.47\textwidth]{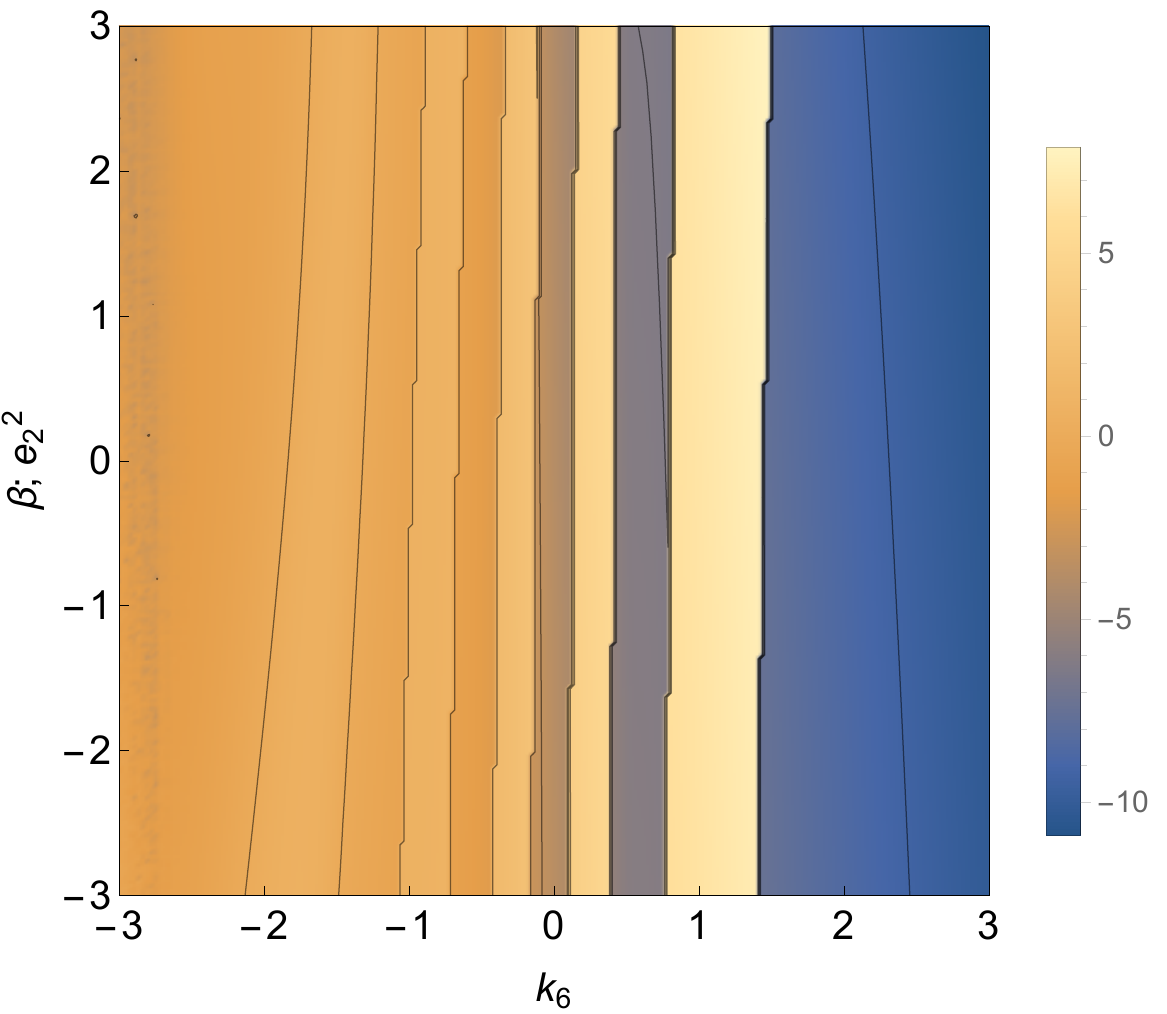}}
\subfigure[]{\label{subfig:ratio3}\includegraphics[width=.47\textwidth]{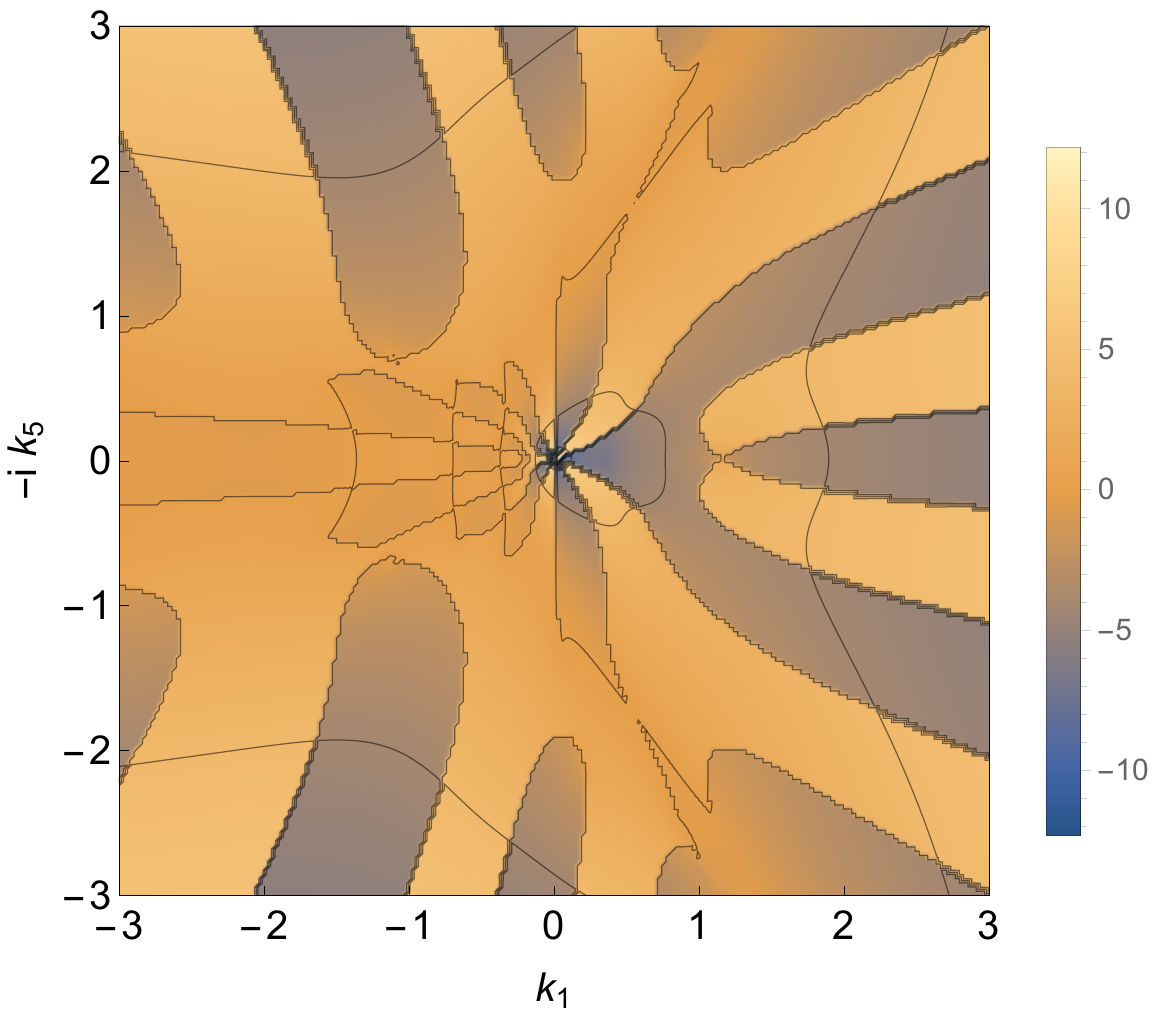}}
\subfigure[]{\label{subfig:ratio4}\includegraphics[width=.47\textwidth]{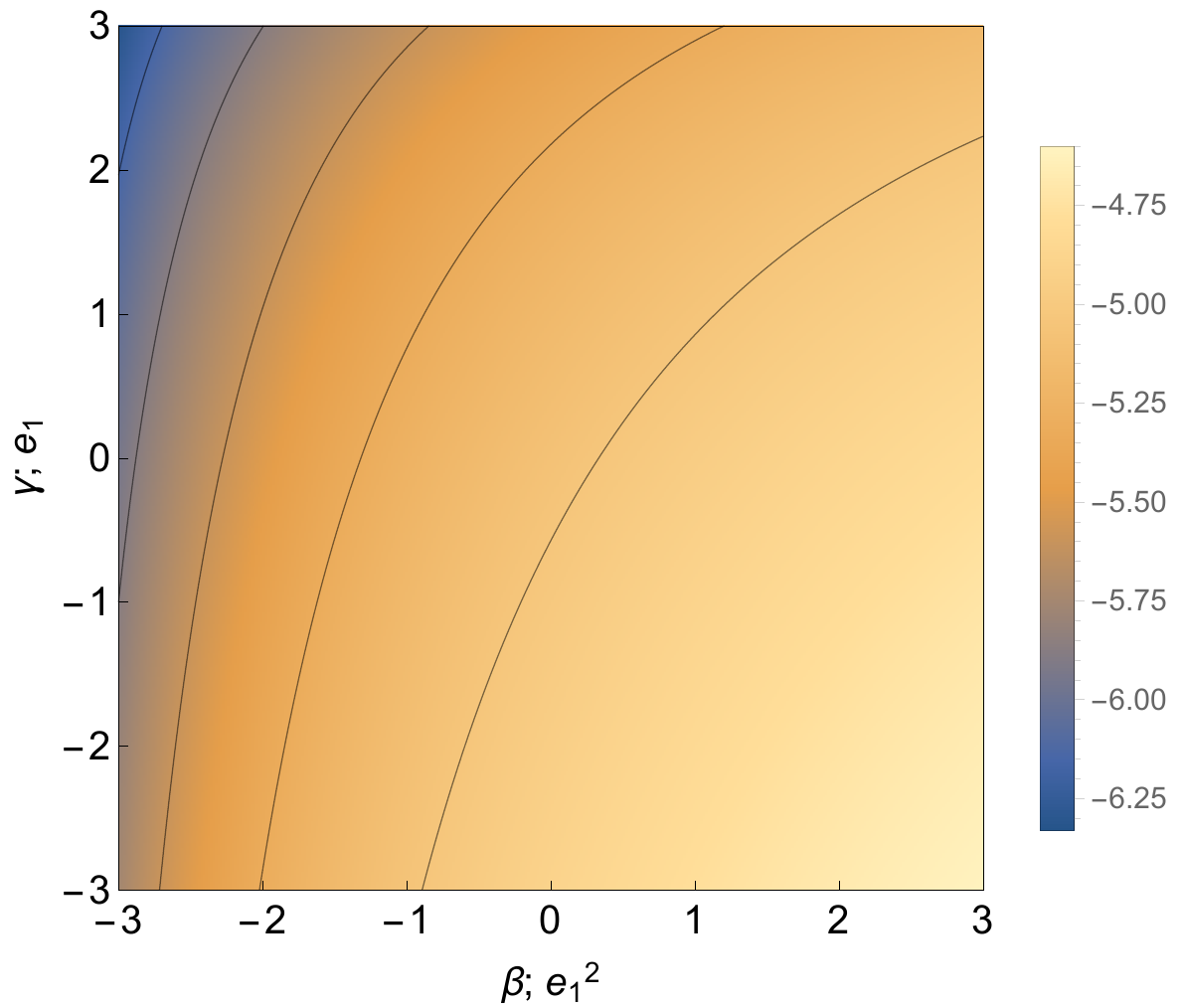}}
\caption{Dependence of Yukawa eigenvalues' ratio $\text{log} |r|$ on the complex structure moduli.
While we use the labels $k_i$ for the parameters of $\delta$ and $d_3$ introduced in \eqref{eq:explicit_param_d3delta}, we indicate the others by the corresponding monomial in the polynomials ($\beta$ or $\gamma$).
In, (a) and (b), we vary one modulus in $d_3$ / $\delta$ and one in $\beta$ / $\gamma$.
In (c), we vary parameters in $d_3$ / $\delta$ only, effectively just moving around the two Yukawa points.
For (d), we only vary parameters in $\beta$ and $\gamma$. 
The results suggest that varying the parameters controlling the Yukawa points affect the ratio more drastically than those in $\beta$ and $\gamma$, unless we modify the coefficients of the high-degree terms in the latter.
These plots were generated in \texttt{Mathematica} and suffer from some numerical instabilities, which do not qualitatively change our results.
}
\label{fig:yukawaRatio}
\end{figure}

To visualize the moduli dependence, we analyze the ratio $r = | \lambda_1 / \lambda_2 |$ for two varying complex structure parameters, while we hold all others fixed at random order 1 values.
It turns out that for order 1 variations, the ratio can develop large hierarchies of ten orders of magnitude, see figure \ref{fig:yukawaRatio}.
In particular, it appears that variations for parameters in the polynomials $d_3$ and $\delta$ affect the ratio more severely than for most parameters in $\beta$ or $\gamma$.
For these parameters, changes by orders of magnitude in $r$ occurs only when we vary the coefficients of the highest degree monomials in $\beta$ or $\gamma$, see figure \ref{subfig:ratio1}.
This observation confirms that the large variations of the relative coupling indeed comes from the prefactor $\gamma\beta^2$ in \eqref{eq:yukawa_final_final}.
Importantly, we find that hierarchies of order $10^3$ and larger are not constrained to lower dimensional subspaces of the complex structure moduli space, but rather generic.
That is, for every pair of parameters we vary, there is finite \textit{area} (rather than just along a line), where we observe such hierarchies.

We emphasize here again that this is only the holomorphic coupling, and the physical values will depend on the flux data and, importantly, the K\"ahler moduli.
However, expectation from earlier works, and also the fact that the observed hierarchies are generic in complex structure moduli, suggest that these additional effects will not affect the holomorphic coupling's hierarchy too much.

We believe that these observations are not a special feature of our model, but rather general for compact F-theory models.
For instance, the rank enhancement can be traced in our step-by-step derivation to the fact that for different basis functions $h^i_{{\bf R}}$ of the wave function zero modes, the rational functions
\begin{align}
  \frac{h^{i_1}_{{\bf R}_1} \, h^{i_2}_{{\bf R}_2}}{f_{{\bf R}_1} \, f_{{\bf R}_2}}
\end{align}
appearing in the residue formula \eqref{eq:yukawa_final_final} have different pole structures at different points with $f_{{\bf R}_1} = f_{{\bf R}_2} =0$.
Such a behavior is expected for general rational functions of this type and therefore clearly not special to our toy model.
The large hierarchies are due to the factor $\gamma\beta^2$ in \eqref{eq:yukawa_final_final}: 
These polynomials have no zeroes at the Yukawa points and thus contribute to the couplings basically as a prefactor given by their values at the points.
However, because they are of rather high degrees ($\beta,\gamma$ have monomials up to degrees $6 / 8$ in $e_i$), these values change by a few orders of magnitudes at the different points.

While the high degree of these polynomials are a direct result of the model, it is not inconceivable that such factors appear also in other examples.
Making this claim on solid footing will require future work.

\section{Conclusions and Outlook}\label{sec:conclusions}

In this work we have demonstrated that global F-theory models can in general exhibit higher rank Yukawa coupling matrices.
At the level of the holomorphic couplings, our analysis has further shown that there are large hierarchies for generic complex structure moduli.
Compared to previous work \cite{Heckman:2008qa,Heckman:2009mn,Hayashi:2009ge,Randall:2009dw,Font:2009gq,Cecotti:2009zf,Conlon:2009qq,Hayashi:2009bt,Marchesano:2009rz,Cecotti:2010bp,Chiou:2011js,Aparicio:2011jx,Font:2012wq,Font:2013ida,Marchesano:2015dfa,Carta:2015eoh}, the key ingredient to our approach is that contributions to the same couplings from \textit{different} Yukawa points in the geometry are in general linearly independent.
In particular, this comes from purely geometric considerations, and does not invoke any instanton or T-brane effects.

For concreteness, we have considered the ${\bf 10}_{-2} \, \bar{\bf 5}_6 \, \bar{\bf 5}_{-4}$ coupling in a compact toy $SU(5) \times U(1)$-model with a $G_4$-flux that induced enough chiral matter to facilitate a higher rank coupling matrix.
On the $SU(5)$-divisor $S_\text{GUT} \cong \text{dP}_2$, we could explicitly parametrize a basis for the wave function zero modes in terms of $\text{dP}_2$ coordinates, because all participating matter curves were rational curves that intersected twice inside a $\mathbb{C}^2$ patch of $S_\text{GUT}$.
Evaluating the corresponding residue formula then became an easy algebra exercise, which indeed confirmed that the two contributions added up to a rank two coupling matrix.

Interestingly, a numerical analysis of the complex structure dependence of the couplings revealed that there is generically a large hierarchy of ${\cal O}(10^{10})$ and more between the two independent holomorphic couplings.
Here, ``generic'' means that we observed these hierarchies in a full-dimensional subspace of the complex structure moduli space.
In our toy model, the origin of these hierarchies can be traced to the factor $\gamma \beta^2$ in the residue formula \eqref{eq:yukawa_final_final} which, since it generically does not share zeroes with the denominator, simply multiplies the value of the residues at the Yukawa points.
However, as a polynomial of degree 20, its value at the different Yukawa points can easily change over several orders of magnitude even if the points are separated by order one changes of the coordinates.
From \eqref{eq:generic_solution_3+2_factorization}, we expect that the degree of $\gamma \beta^2$ is generally very high, since it appears as factor of $a_{6,5}$ which itself (as section of ${\cal O}_B(6\overline{K}_B - 5 S_\text{GUT})$) is typically a high degree polynomial.
This does not rule out models, particularly of other fibration type with different spectral cover descriptions, where the relevant polynomials are of low degree and thus might have a less prominent hierarchy at generic values of complex structure.
Whether such models are easily constructed or, perhaps more importantly, can exhibit other phenomenologically appealing aspects, will hopefully be answered in future works.

Note that so far, we have only discussed the holomorphic Yukawa matrices.
While we would need to also compute the K\"ahler-moduli dependent normalization factors of the wave functions to obtain the physical couplings, the expectation---also based on intuition from type II compactifications \cite{Cvetic:2003ch,Cremades:2003qj,Cremades:2004wa,Blumenhagen:2006xt,Blumenhagen:2007zk}---is that these factors do not affect the hierarchies strongly.
In particular, given that these are generic in complex structure moduli, it would be highly unlikely if these non-holomorphic factors always conspire to cancel the K\"ahler-moduli independent hierarchy of the holomorphic couplings.

It would clearly be interesting to adapt our computation to models with more phenomenological appeal than our toy model.
In particular, demonstrating in the recently found class of three-family MSSM models \cite{Cvetic:2019gnh} that the up-type quark mass matrix generically has rank three with large hierarchies---even just at the level of holomorphic couplings---could provide a strong argument for ``string universality'' in the particle physics sector of F-theory.

To achieve this, there is clearly more technical and conceptual details to be understood.
For one, finding an explicit parametrization on the gauge divisor of holomorphic sections on higher genus curves will require more elaborate techniques than for $\bbP^1$s.
More importantly, it will be challenging to find an appropriate map between the Higgs bundle description and the global geometry in cases without a known spectral cover description.
And finally, it will be imperative to also understand the non-holomorphic prefactors encoding the K\"ahler moduli dependence in a global setup, in order to determine the physical couplings.
We look forward to address these issues in future works.

\section*{Acknowledgments}

We thank Jonathan Heckman and Craig Lawrie for useful discussions.
The work of MC and LL is supported by DOE Award DE-SC0013528Y.
MC further acknowledges support from the Slovenian Research Agency No.~P1-0306, and the Fay R.~and Eugene L.~Langberg Chair funds.
The work of GZ is supported by NSF CAREER grant PHY-1756996.

\appendix

\section{Details on the \texorpdfstring{dP\boldmath{${_2}$}}{dP2} Geometry}\label{app:toric_dP2}

\begin{figure}[h]
  \centering
  \begin{tikzpicture}
    \filldraw[black] (0,0) circle (0.5pt);
    \filldraw[black] (0,2) circle (2pt);
    \filldraw[black] (2,0) circle (2pt);
    \filldraw[black] (-2,0) circle (2pt);
    \filldraw[black] (-2,2) circle (2pt);
    \filldraw[black] (0,-2) circle (2pt);
    \draw[gray, thick] (0,0) -- (0,2);
    \draw[gray, thick] (0,0) -- (0,-2);
    \draw[gray, thick] (0,0) -- (2,0);
    \draw[gray, thick] (0,0) -- (-2,2);
    \draw[gray, thick] (0,0) -- (-2,0);
    \draw[gray, thick] (-2,2) -- (0,2);
    \draw[gray, thick] (-2,2) -- (-2,0);
    \draw[gray, thick] (0,2) -- (2,0);
    \draw[gray, thick] (2,0) -- (0,-2);
    \draw[gray, thick] (0,-2) -- (-2,0);
    \filldraw[black] (-2,2) circle (2pt) node[anchor=south east] {$u$};
    \filldraw[black] (0,2) circle (2pt) node[anchor=south] {$e_1$};
    \filldraw[black] (2,0) circle (2pt) node[anchor=west] {$w$};
    \filldraw[black] (0,-2) circle (2pt) node[anchor=north] {$v$};
    \filldraw[black] (-2,0) circle (2pt) node[anchor=east] {$e_2$};
  \end{tikzpicture}\caption{The toric diagram of dP$_2$}
  \label{fig:toric_diagram_dP2}
\end{figure}
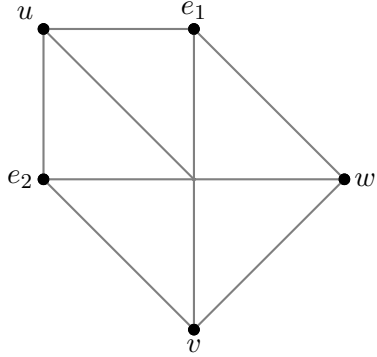

In this appendix we provide some useful details about the geometry.
On dP$_2$ one can introduce toric coordinates $(u,v,w,e_1,e_2)$, which are sections of the following line bundles:
\begin{equation}
  [u] = H-E_1-E_2 \, , \quad  [v] = H-E_2 \, , \quad [w]=H-E_1 \, , \quad [e_1] = E_1 \, , \quad [e_2]=E_2 \, .
\end{equation}
Given any line bundle on $S_\text{GUT}$, we can write a generic section of it as homogeneous polynomials in these coordinates.
The Stanley--Reisner generators of dP$_2$, that is, combinations of toric variables which cannot vanish simultaneously, is given by
\begin{align}\label{eq:SRI_dP2}
  \{ w e_2,  w u, v e_1 , e_2 e_1, v u \}\, .
\end{align}
This information can be extracted from a reflexive polygon, the toric diagram of dP$_2$, which for completeness we present in figure \ref{fig:toric_diagram_dP2}.

The homology class of an irreducible curve $C \subset \text{dP}_2$ can be written as
\begin{equation}
[C] = n_H \, H + n_{E_1} \, E_1 +n_{E_2} \, E_2, \quad (n_H \geq 0 \, , \, n_{E_i} \leq 0) \, .
\end{equation}
Note that if $n_{E_i} > 0$, $C$ is not irreducible as there is always a factor of the blow-up curve $\{e_i\}$ with some multiplicity appearing.

For the polynomials $d_3$ and $\delta$ parametrized as \eqref{eq:explicit_param_d3delta}, we can derive that the intersection points $d_3 =0 =\delta$ must be in the patch with $u,v,w \neq 0$.
For example, setting $u=0$ the equation for $d_3$ yields $k_4 v w$, which cannot be zero since both $v$ and $w$ are in the Stanley--Reisner ideal \eqref{eq:SRI_dP2} with $u$; thus $u$ cannot be 0 when $d_3$ vanishes.
A more practical way to argue for it is to simply check that in the patch $(e_1,e_2)$ (i.e., when we set $u=v=w=1$), there are two distinct solutions to $\delta=0=d_3$:
\begin{align}\label{eq:solutions_yukawa_points}
  \begin{split}
    e_1 & = \frac{k_0\, k_4 - k_2\, k_5 + k_1\, k_6 \pm \sqrt{4\, k_1\, k_4\, (k_2\, k_3 - k_0\, k_6) + (k_0\, k_4 - k_2\, k_5 + k_1\, k_6)^2}}{2\,(k_2\,k_3 - k_0\,k_6)} \, ,\\
    e_2 & = \frac{k_0\, k_4 + k_2\, k_5 - k_1\, k_6 \mp \sqrt{4\, k_1\, k_4\, (k_2\, k_3 - k_0\, k_6) + (k_0\, k_4 - k_2\, k_5 + k_1\, k_6)^2}}{2(k_1 \, k_3 - k_0\, k_5)} \, .
  \end{split}
\end{align}
Since we know that $[d_3]\cdot [\delta]=2$, these must be all intersections, which indeed lie in the claimed $\mathbb{C}^2$ patch of dP$_2$.

\bibliography{references}{}
\bibliographystyle{JHEP} 

\end{document}